
\input phyzzx
\overfullrule=0pt
\def\IR{\relax{\rm I\kern-.18em R}}
\tolerance=5000
\overfullrule=0pt
\font\big=cmbx12 scaled\magstep2

\pubnum{IASSNS-HEP-91/53}
\date{August, 1991}
\centerline{\big ON BLACK HOLES IN STRING THEORY\foot{Lecture
at Strings '91, Stonybrook, June 1991.}}
\bigskip
\centerline{Edward Witten\foot{Research supported in part by NSF Grant
PHY86-20266.}}
\medskip
\centerline{\sl School of Natural Sciences}
\centerline{\sl Institute for Advanced Study}
\centerline{\sl  Olden Lane}
\centerline{\sl Princeton, NJ 08540}
\bigskip\bigskip
\REF\mtw{C. W. Misner, K. S. Thorne, and J. A. Wheeler, {\it Gravitation}
(W. H. Freeman and Co., 1973), especially pp. 831-2.}
In its most elementary form, the Schwarzschild solution is
$$ds^2=-\left(1-{2GM\over r}\right) dt^2+\left(1-{2GM\over r}\right)^{-1}
dr^2
+r^2 d\Omega^2. \eqn\joffo$$
One quickly sees that the space-time described by this metric is not
geodesically complete and hence that there must be more to the story.
Indeed, analytic continuation of the Schwarzschild solution,
as described for instance in [\mtw],
transforms it to a form
$$ds^2=-{du\,\,\,dv\over 1-uv}e^{w(u,v)}+r(u,v)^2d\Omega^2, \eqn\offo$$
 from which the global properties can be understood.
Here $r(u,v)$ and $w(u,v)$ are certain functions whose details are
not so material.  The important properties come from the singularities
at $uv=1$.  There are two branches of this singularity, as $u$ and $v$
may be both negative or both positive.   The ``physical region'' is the
region $uv<1$ between the two singularities.
The branch of the singularity with $u,v<0$ is
the ``white hole'' singularity; it is to the past of the physical region,
and can emit but not absorb matter.  The branch of the singularity with
$u,v>0$ is the ``black hole'' singularity; being to the future of the
physical region, it can absorb but not emit.
The white hole violates the predictability of the classical theory,
as classically one cannot predict what it will emit.  Relativists
have dealt with the white hole by the {\it cosmic censorship hypothesis},
according to which white holes and more general ``naked singularities''
never form from acceptable initial data.
(For instance, spherical stellar collapse from standard initial data
gives a spacetime that coincides
with the Schwarzschild solution only in the {\it exterior} of the
collapsing star;
the exterior contains the black hole singularity but not the white
hole singularity.)  Actually, there is not very much evidence for the
cosmic censorship
conjecture, and it may well be false.  It is not clear that we should
wish for the truth of the conjecture.  If cosmic censorship is false,
then in principle we would have the chance to observe
new laws of physics that must take over near the would-be naked singularity.
This might be more useful for physics than extending the scope of classical
general relativity by proving cosmic censorship.

It is usually assumed that the new physical laws associated with a possible
breakdown of cosmic censorship would involve quantum gravity,
but this may be wrong, particularly in view of the fact that the
length scale of string physics is probably a little larger than the
Planck length.
If indeed cosmic censorship is false in general relativity
but its analog is true in string theory, then it may well be that
it is classical string theory that becomes manifest near the would-be
naked singularity.

As for black holes, at the classical level they cause no breakdown in
predictability for the {\it outside observer}.  Quantum mechanically,
though, the idea that black holes exist and white holes do not is
paradoxical as the white hole is the CPT conjugate of the black hole.
In any event, Stephen Hawking rendered the classical picture obsolete
with his discovery of black hole radiation, showing that an isolated
hole will {\it radiate} until (after Hawking's approximations break down)
it reaches a quantum ground state or disappears completely.
A key aspect of the problem of {\it quantum mechanics of black holes} is to
describe
this endpoint.  More broadly, one would like to describe the $S$-matrix for
matter interacting with the black hole if (as I suspect)
there is one.  If (as argued
by Hawking) such an $S$-matrix does not exist, one would like to
get a precise account of the nature of the obstruction and to learn
how to calculate in whatever framework (such as the density matrices
advocated by Hawking) replaces that of $S$-matrices.

I think most physicists expect that Hawking radiation leads to
complete disappearance of neutral black holes.
For charged black holes the situation is likely to be quite different,
under appropriate conditions, since a charged hole might be lighter
than any collection of ``ordinary particles'' of the same total charge.
Let us call the electric and magnetic charges of the hole $q$ and $m$.
A charged black hole has a {\it classical ground state} with mass
$M=\sqrt{q^2+m^2}$ in Planck units; this ground
state is simply the extreme Reissner-Nordstrom solution of the classical
field equations.  One would imagine that for suitable values of
the charges the classical ground state is some sort of approximation to
the quantum ground state.

In this respect, the case of a magnetically charged black hole is
especially interesting:

\REF\ufus{S. Orito et. al., ``Search For Supermassive Relics With a
$2000{\rm m}^2$ Array Of Plastic Track Detectors,'' Phys. Rev. Lett.
{\bf 66} (1991) 1951.}
(1) This case may be realized in nature
if $M_{\mit GUT}\sim M_{\mit Planck}$, since
then the 't Hooft-Polyakov monopole is a black hole.
In particular it is at least conceivable that the dark matter in our galactic
halo could consist of magnetically charged black holes.  This hypothesis
is subject to experimental test; indeed recent experiments are relatively
close to the sensitivity required to exclude it [\ufus].
Any type of discovery of galactic halo particles would give particle physics
a big boost, of course, but magnetic black holes would be particularly
exciting.

(2) For $e<<1$, the magnetic black hole is much heavier than the Planck mass
and much larger than the Planck radius.  Quantum gravity should therefore,
in a suitable sense, have only a weak effect on the structure of the
magnetic black hole, and such objects might well be accessible to human
understanding, maybe even relatively soon.
In particular, the deviation of the mass of a magnetic black hole from
the classical value is likely to be a quantum gravity or string theory
effect of order $e^2/\hbar c$.  It would certainly be an unusually
interesting thing to measure if magnetic black holes were ever discovered.

(3) For $M_{\mit GUT}<<M_{\mit Planck}$, Callan and Rubakov have already
solved a problem that is a protoype of what we want to do for $M_{\mit GUT}
\sim M_{\mit Planck}$.  They considered the case of a Dirac point monopole
interacting with charged fermions.  This classical system does not quite
make sense, since there are certain classical $s$ wave modes that go in but do
not come out, or go out but did not come in; in more technical terms,
to give a self-adjoint extension of the Hamiltonian requires additional
information not present in the classical theory.  The problem has a striking
analogy with the Schwarzschild space-time where again the problem is that
there are modes (emitted from the past singularity) that go out but
did  not come in, and there are other modes (absorbed by the future
singularity) that go in but do not go out.
In solving the $M_{\mit GUT}<<M_{\mit Planck}$ problem,
Callan and Rubakov showed that a two dimensional $s$-wave effective field
theory gives a good description of monopoles in that regime.
It seems natural to hope that this is still true if $M_{\mit GUT}\sim
M_{\mit Planck}$.  That would mean that some of the important qualitative
aspects of black hole physics could
be described by an effective two dimensional field theory.

In the case studied by Callan and Rubakov, the relevant two dimensional
field theory is weakly coupled, but still requires careful analysis
and exhibits striking phenomena, precisely because a naive form
of the weak coupling limit would give back the pathologies of the classical
system.  It is reasonable to hope that their work is a prototype for
at least some aspects of the black hole problem.

\REF\forge{S. Elitzur, A. Forge, and E. Rabinovici, ``Some Global
Aspects of String Compactifications,'' preprint RI-141(90).}
\REF\bard{K. Bardakci, M. Crescimanno, and E. Rabinovici, ``Parafermions
{}From Coset Models,'' LBL preprint (1990).}
\REF\roc{M. Rocek, K. Schoutens, and A. Sevrin, ``Off-Shell WZW Models
In Extended Superspace,'' IASSNS-HEP-91/14.}
\REF\wad{G. Mandal, A. Sengupta, and S. Wadia, ``Classical Solutions
Of 2-Dimensional String Theory,'' IASSYS-HEP/91/10.}
\REF\witten{E. Witten, ``String Theory And Black Holes,''
Phys. Rev. {\bf D44} (1991) 314.}
What is the status of black holes
in string theory?  One might have believed that, because
of the excellent short distance behavior, there would be no classical
singularities in string theory.  But it has been found recently that
the very metric $ds^2=-du\,\,\,dv/(1-uv)$ that is the essence of the black
hole gives a conformal field theory.  In fact, there is a conformal
field theory with a two dimensional target space, parametrized by two variables
$u$ and $v$, and a
world-sheet action
$$I=\int d^2\sigma\left\{\left({\partial_\alpha u\partial^\alpha v\over 1-uv}
\right)                    +R^{(2)}\Phi(u,v)\right\}, \eqn\odolo$$
where $\Phi(u,v)=\ln(1-uv)+{\rm constant}$ is the dilaton field.
The Euclidean version of this solution was first found by Elitzur,
Forge, and Rabinovici [\forge].  Different forms of the solution
were rediscovered by several
authors from various points of view [\bard,\roc,\wad] (and implicitly
discussed in current algebra by Bars; see his lecture at this conference
and references therein) before I rediscovered
the solution
as an $SL(2,\IR)/U(1)$ coset model and interpreted it as a black hole
[\witten].

It is no accident that this system involves a target space-time
so similar to the Schwarzschild solution.
The general $s$ wave
ansatz of four dimensional general relativity
$$ds^2=g_{\alpha\beta}dx^\alpha dx^\beta+e^\Phi d\Omega^2 \eqn\oco$$
involves a two dimensional metric tensor $g$ and a two dimensional
scalar field $\Phi$.  These are analogous to the metric and dilaton field
of the low energy limit of string theory, and indeed the two dimensional
action that one gets by dimensional reduction of four dimensional general
relativity with the ansatz \oco\ is very similar to the familiar
lowest order action
$$I=\int d^2x \sqrt g e^\Phi\left(R+(\nabla\Phi)^2+8\right)\eqn\doof$$
of the graviton-dilaton system in string theory.  The analogy is even
better if one considers the dimensional reduction of general relativity
in the presence of a spherically symmetric magnetic field, so as to get
a system in which the black hole has a classical ground state.

If one writes the black hole sigma model \odolo\ in Schwarzschild-like
coordinates, and traces what the Hawking radiation ought to mean in this
situation [\witten], one sees, at least heuristically, what the endpoint
of the Hawking radiation ought to be: it should be the standard flat
space-time of the usual $c=1$ model, with a linear dilaton field.
In making an analogy of the $c=1$ model with four dimensional black holes,
this makes it clear how we should think about the linear dilaton field:
it is the ``field'' of an ``object'' sitting at $r=-\infty$ (where the
tachyon potential corresponding to the world-sheet Liouville interaction
is large).
The standard $c=1$ flat space-time should therefore be thought of
as an analog, not of four dimensional Minkowski space, but of
the extreme Reissner-Nordstrom
solution in four dimensions.  This enables us to understand intuitively
the otherwise annoying absence of Poincar\'e invariance in the model.

The field theory of the $s$-wave sector of general relativity is
an interesting two dimensional field theory, which for many years has
looked
like a tempting model of the quantum physics of black holes -- particularly
in view of the analogy with the Callan-Rubakov effect that I explained before.
Of course to get a real model of black
hole physics, one must couple this system to ``matter.''  The $s$-wave
component of a neutral scalar field would be a satisfactory form of matter
mathematically, but there is a perhaps more
``physical'' candidate -- in the field of a  monopole of minimal Dirac
charge, the usual quarks and leptons have $s$-wave components (which
are important in the Callan-Rubakov effect).
The combined  system of spherically symmetric geometry and $s$-wave fermions
is weakly coupled if $e<<1$.
Despite the weak coupling, the  $s$-wave field theory of magnetic
black holes has defied understanding (at least at my hands).

String theory with $c=1$ or $D=2$ is superficially a very similar system,
with the massless ``tachyon'' playing the role of the bosonized
$s$-wave fermions.
It is believed to be exactly soluble via matrix models.  The already
computed $S$-matrix of this model can probably be understood as a long-sought
example of an $S$-matrix for matter interacting quantum mechanically
with a black hole.
If we could get a good understanding of the
space-time interpretation of the matrix model results, we could probably
sharpen our understanding of black holes.

The most striking puzzle in this area is
probably the apparent absence in the matrix model of an analog of the
expected  back-reaction
of matter on the gravitational field.
In other words, if $D=2$ string theory really does describe target
space gravity, one would expect to see in the theory a suitable analog
of the back-reaction of matter on gravity.  In the matrix model,
all that we presently see that might even loosely be regarded as
``gravity'' is the one body Hamiltonian of the free fermions (which is the
``background'' in which they are moving).  But  the free fermions have
no back reaction on their own Hamiltonian.  Perhaps some degrees of freedom
(analogous to the dilaton?) have been suppressed or integrated out in the
matrix model description.  A proper understanding of
this point will probably help make it clear how much can be learned about
real black holes from $D=2$ \break string theory.

\refout
\end